\documentclass[useAMS,twocolumn]{mn2e}
\usepackage{graphicx}
\usepackage{savesym}
\usepackage{listings}
\usepackage[intlimits]{amsmath}%option changes placement of integral limits
\savesymbol{iint}
\usepackage{txfonts}
\usepackage{bm}
\restoresymbol{TXF}{iint}
\usepackage{amssymb}
\usepackage{amsmath}
\usepackage{tabularx}
\usepackage{color}
\usepackage{multicol,lipsum}

\usepackage{comment}

\def\mnras{Mon. Not. Roy. Astron. Soc. }

\def\gsim{\mathrel{\raise.5ex\hbox{$>$}\mkern-14mu
             \lower0.6ex\hbox{$\sim$}}}
\def\lsim{\mathrel{\raise.3ex\hbox{$<$}\mkern-14mu
             \lower0.6ex\hbox{$\sim$}}}

\title[EM GWs from Pulsars]{Gravitational Waves from the Pulsar Magnetosphere}% Force line breaks with \\
\author[ Contopoulos, Kazanas, Papadopoulos 2023]{{Ioannis Contopoulos$^{1}$\thanks{Email: icontop@academyofathens.gr}, Demosthenes Kazanas$^{2}$\thanks{Email: demos.kazanas@nasa.gov}, Demetrios B. Papadopoulos$^{3}$\thanks{Email:papadop@astro.auth.gr}} %\vspace{0.4cm}
\\
\parbox{\textwidth}{$^{1}$ Research Center for Astronomy and Applied Mathematics, Academy of Athens, 4 Soranou Efessiou Str., Athens 11527, Greece}\\
\parbox{\textwidth}{$^{2}$ Astrophysics Science Division, NASA/ Goddard Space Flight Center, Greenbelt Maryland, 20771, USA}\\
\parbox{\textwidth}{$^{3}$ Aristotle University of Thessaloniki, Department of Physics, Section of Astrophysics, Astronomy and Theoretical Mechanics, Thessaloniki  54124, Greece}}

\begin{document}

\date{Accepted -. Received -; in original form -}
\pagerange{\pageref{firstpage}--\pageref{lastpage}} \pubyear{2023}
\maketitle

\label{firstpage}

\begin{abstract}
We investigate the generation of gravitational waves from the rotation of an orthogonal pulsar magnetosphere in flat space time. We calculate the first order metric perturbation due to the rotation of the non-axisymmetric distribution of electromagnetic energy density around the central star. We show that gravitational waves from a strong magnetic field pulsar right after its formation within a distance of 1~kpc may be detectable with the new generation of gravitational wave detectors.

\end{abstract}

%\pacs{Valid PACS appear here}
\maketitle

%\tableofcontents

\section{Introduction}

Gravitational waves are ripples in the fabric of spacetime caused by the acceleration of massive objects \cite{Rainich1925,Misner1973,Ivanov2008}.
In particular, rotating neutron stars, the remnants of massive stars that have undergone supernova explosions, are potential sources of continuous gravitational wave emission when irregularities or deformations of the star's spherical form are present. As the star's rotation rate remains steady, the emitted gravitational waves will possess consistent frequency and amplitude, akin to a singer holding a sustained note. The study of these emissions, if detected, can provide valuable insights into the properties of neutron stars. Rotating neutron stars are also strongly magnetized, and are thus surrounded by a magnetosphere that is known to emit electromagnetic (Poynting) radiation, and is often observed to also emit periodic electromagnetic pulses at the period of their rotation (milliseconds to tens of seconds).

In the present work we will show that a non-axisymmetric rotating magnetosphere can also emit gravitational waves. We will ignore non-spherical deformations of the neutron star, and we will investigate continuous gravitational waves that are due solely to the imprint of the anisotropic rotating distribution of electromagnetic energy on the metric. As a first approximation, we will ignore the effect of the star's own gravity and will consider the gravitational perturbation of a rotating electromagnetic field on flat space time. Notice that already in 1925, Rainich gave the necessary and sufficient conditions for a gravitational field to originate from a no-null electromagnetic field

%\section{The perturbed Einstein Field Equations.}

We will assume that the metric  $g_{\mu\nu}$ can be expanded
 in powers of a small parameter $\varepsilon \ll 1$
\begin{equation}\label{e1}
g_{\mu\nu}^{(exact)}=g_{\mu\nu}+\varepsilon h_{\mu\nu}+O(\varepsilon^2)
\end{equation}
%where $h_{\mu\nu}$ %(denoted also as $h$)
%is the total space time perturbation
(see~\cite{Thorne1982,Wald1984,Maggiore2008,Bambi2022})
and the stress-energy tensor  $T_{\mu\nu}$ can be similarly expanded as
\begin{equation}\label{e2}
T_{\mu\nu}=\varepsilon T_{\mu\nu}^{(1)}+O(\varepsilon^2)
\end{equation}
(Greek indices take values $0,1,2,3$, Latin indices $1,2,3$).
To expand the Einstein equations $G_{\mu\nu}=8\pi T_{\mu\nu}$ in powers of $\varepsilon$, we expand the Einstein tensor in powers of the perturbation $h$
\begin{equation}\label{e3}
G_{\mu\nu}(g+h)=G_{\mu\nu}(g)+ G_{\mu\nu}^{(1)}(h)+O(h^2)
\end{equation}
where, $G_{\mu\nu}^{(1)}$ is obtained from the exact Riemann tensor~\cite{Wald1984}
(see Ch.4, page 74). For a vacuum background keeping only the first order terms in our approximations we obtain
\begin{equation}\label{e4}
G_{\mu\nu}^{(1)}=(g_{\mu}^{\alpha} g_{\nu}^{\beta}-\frac{1}{2} g_{\mu\nu} g^{\alpha\beta})R_{\alpha\beta}^{(1)}
\end{equation}
where
\begin{equation}\label{e6}
R_{\alpha\beta}^{(1)}=-\frac{1}{2}(\Box h_{\alpha\beta}+2g^{\mu\lambda}g^{\nu\kappa}R_{\alpha\lambda\beta\kappa} h_{\mu\nu}-2g^{\mu\sigma}\bar{h}_{\mu
(\alpha;\sigma\beta)})
\end{equation}
Here, $\Box \equiv g^{\mu\nu}\nabla_{\mu}\nabla_{\nu}$ is the d'Alembertian in any background and
\begin{equation}\label{e8}
\bar{h}_{\mu\nu}\equiv h_{\mu\nu}-\frac{1}{2}g_{\mu\nu} g^{\alpha\beta} h_{\alpha\beta}
\end{equation}
A semicolon and $\nabla$ both denote the covariant derivative, compatible with the flat space time $g_{\mu\nu}$.
Noting that $\bar{h}=g^{\mu\nu}\bar{h}_{\mu\nu}=h-2h=-h$, eq.~(\ref{e8}) can be inverted to give
\begin{equation}\label{e8b}
h_{\mu\nu}=\bar{h}_{\mu\nu}-\frac{1}{2}g_{\mu\nu}\bar{h}
\end{equation}
Substitution of eqs.~(\ref{e1}) and (\ref{e2}) into Einstein equations and equating only powers of $\varepsilon$, we %keep only the term
end up with the equation
\begin{equation}\label{e9}
G_{\mu\nu}^{(1)}=8\pi T_{\mu\nu}^{(1)}
\end{equation}
This perturbative expansion comes with the freedom to perform gauge transformations
\begin{equation}\label{e11}
h_{\mu\nu}\rightarrow h_{\mu\nu}+{\cal L}_{\xi} g_{\mu\nu}
\end{equation}
where ${\cal L}_{\xi}$ is a Lie derivative, and $\xi^{\alpha}$ are the components of a freely chosen vector field.

In self-force theory, this freedom is commonly used to impose the
Lorenz gauge condition
\begin{equation}\label{e13}
\nabla_{\alpha}\bar{h}^{\alpha\beta}=0
\end{equation}
in which case the linearized Einstein tensor simplifies to
\begin{equation}\label{e14}
G_{\mu\nu}^{(1)}=-\frac{1}{2}(\Box\bar{h}_{\mu\nu}+2g^{\alpha k}g^{\lambda \beta}R_{\mu\kappa\nu\lambda}\bar{h}_{\alpha\beta})
\end{equation}
eq.~(\ref{e9}) with the aid of eqs.~(\ref{e8}) and (\ref{e14}) gives
\begin{equation}\label{e15}
-\frac{1}{2}(\Box\bar{h}_{\mu\nu}+2g^{\alpha k}g^{\lambda \beta}R_{\mu\kappa\nu\lambda}\bar{h}_{\alpha\beta})=8\pi T_{\mu\nu}^{(1)}
\end{equation}

We will solve eq.~(\ref{e15}) in a $(t,r,\theta,\phi)$ coordinate system in a Minkowski background where  $g_{\mu\nu}=\eta_{\mu\nu}$. To do so, we choose $\xi$ such that the trace $\bar{h}=0$ and thus $h_{\mu\nu}=\bar{h}_{\mu\nu}$~\cite{Maggiore2008}.
%; Ch.~1, page~8.
Next, we will consider the metric perturbations $h_{\mu\nu}$ that are due to a rotating non-axisymmetric strong electromagnetic field.

{
We recently came across the most interesting paper Nazari \& Roshan~(2020). These authors investigate the same problem, their method is, however, different. They implement a post-Newtonian (PN) approximation to the energy–momentum tensor of the magnetized fluid, and a PN expansion of the gravitational potential in the wave zone. We instead work directly with the perturbation of the stress-energy tensor and study its effect on the background metric. We will compare our respective results in the Discussion section.
}

%####################################################################################
\section{A Rotating Magnetosphere}
The Minkowski space time in spherical coordinate system $(t,r,\theta,\phi)$ in geometric units where $G=c=1$ is
\begin{equation}\label{p0}
g_{\mu\nu}=\eta_{\mu\nu}=diag(1,-1,-r^2,-r^2\sin^2{\theta})
\end{equation}
and eq.~(\ref{e15}) reads~\cite{Wald1984,Maggiore2008,Bambi2022}
\begin{equation}\label{p0a}
\Box h_{\mu\nu}=-16\pi T_{\mu\nu}^{(1)}\ .
\end{equation}
%where $h_{\mu\nu}\equiv h_{\mu\nu}^{(1)}$.

We will assume that the energy-momentum tensor is due entirely to an electromagneti field and is thus given by
\begin{equation}
T^{(1)}_{\mu\nu}=-\frac{1}{4\pi}(F_{\mu\sigma} F_{\nu\rho}g^{\rho\sigma}-\frac{1}{4}F_{\rho\sigma} F^{\rho\sigma}g_{\mu\nu} )\ ,
\label{p0b}
\end{equation}
where the Faraday tensor $F^{\mu\nu}$ can be constructed from the spatial vectors ${\bf E}$, ${\bf B}$ of the electric and magnetic field.

The electromagnetic field that we will investigate is the central magnetic dipole of an orthogonally rotating neutron star, otherwise known as an orthogonal pulsar. The rotating spherical source at the origin has size $r_0\sim 10\ {\rm km}$ and angular velocity of rotation $\omega_0\equiv 2\pi/P$, where $P$ is the period of rotation. We will only consider the stellar exterior, assuming that the mass and magnetic field distribution in the stellar interior is more or less spherically symmetric and does not generate gravitational waves. We will consider a rotating dipole field of the form
\begin{equation}
{\bf B} = \frac{3\hat{\bf r}(\hat{\bf r}\cdot {\bf m})-{\bf m}}{r^3}
\end{equation}
where ${\bf m}=m(\cos({\omega_0 t}),\sin{(\omega_0 t)},0)$ in cartesian coordinates $(x,y,z)$, $z$ is the axis of rotation, and $\hat{\bf r}$ is the unit vector along the radial direction that connects a point in space with the origin. Here, $m\equiv B_0 r_0^3$ is the stellar magnetic dipole, and $B_0$ is the equatorial value of the stellar magnetic field. In this configuration, the covariant components of the magnetic field are
\begin{eqnarray}
B^r &=&  2B_0\frac{r_0^3}{r^3}\sin\theta\ \cos(\omega_0 t -\phi)
\label{Br}\\
B^\theta &=& -B_0\frac{r_0^3}{r^3}\cos\theta\ \cos(\omega_0 t -\phi)\\
B^\phi &=& -B_0\frac{r_0^3}{r^3}\sin(\omega_0 t -\phi)\ ,
\label{Bphi}
\end{eqnarray}
Assuming ideal conditions in the magnetosphere, it is straightforward to show (see Appendix~A) that, in steady state magnetospheric rotation,
\begin{equation}
{\bf E} = -\omega_0 r\sin\theta\ \hat{\phi}\times {\bf B}\ ,
\label{Eideal}
\end{equation}
which yields
\begin{eqnarray}
E^r &=& -\frac{B_0 r_0^3\omega_0}{r^2}\sin\theta\cos\theta\ \cos(\omega_0 t -\phi)
\label{Er}\\
E^\theta &=& -\frac{2B_0 r_0^3\omega_0}{r^2}\sin^2\theta\ \cos(\omega_0 t -\phi)\\
E^\phi &=& 0\ .
\label{Ephi}
\end{eqnarray}
Here, $\hat{\phi}$ is the unit vector in the azimuthal direction. The reader can directly check that the above sets of eqs. ({\ref{Br}-\ref{Bphi}) and (\ref{Er}-\ref{Ephi}) satisfy the ideal Maxwell's equations which in geometric units read
\begin{eqnarray}
&&\nabla \cdot {\bf E}=4\pi \rho_e\ \ \nabla \cdot {\bf B}=0
\nonumber\\
&&\label{em2a}
\frac{\partial {\bf E}}{\partial t}=\nabla\times {\bf B}-4\pi {\bf J}\approx 0\ \ \frac{\partial {\bf B}}{\partial t}=\nabla\times {\bf E}\approx 0\ .
%\label{Maxwell}
\end{eqnarray}
%with the force-free condition
%\begin{equation}\label{forcefree}
%\rho_e{\bf E}+{\bf J}\times {\bf B}=0\ .
%\end{equation}
Here $\rho_e$ is the electric charge density, and ${\bf J}$ is the magnetospheric electric current.
% which satisfies the charge conservation  equation
%\begin{equation}\label{em2b}
%\frac{\partial\rho_e}{\partial t}+\nabla\cdot{\bf J}=0
%\end{equation}

From the study of pulsar magnetospheres it is known that the above dipole structure of the magnetic field changes significantly beyond the cylindrical distance of the so-called light cylinder, namely $R_{\rm lc}\equiv c/\omega_0$. It is interesting that, beyond that distance, the magnetic field attains a configuration that may be described as a `wound monopole', with
\begin{eqnarray}
|B^r|(r\sin\theta\gsim R_{\rm lc}) &\propto & \frac{1}{r^2}
\label{Brfar}\\
B^\theta(r\sin\theta\gsim R_{\rm lc}) &\approx& 0\\
|B^\phi |(r\sin\theta\gsim R_{\rm lc})&\propto & |B^r|\frac{r\sin\theta}{R_{\rm lc}} \propto \frac{\sin\theta}{r}
\label{Bphifar}
\end{eqnarray}
\cite{Bogovalov1999}.
It is obvious from the above eqs.~(\ref{Brfar}-\ref{Bphifar}) that beyond the light cylinder, the distribution of $B^2$ is axisymmetric to a good approximation (i.e. it does not depend on the azimuthal angle $\phi$), and is, therefore, incapable of generating gravitation radiation. In contrast, the distribution of the magnetic and electric field well inside the light cylinder as expressed via eqs.~(\ref{Br}-\ref{Bphi}) and eqs.~(\ref{Er}-\ref{Ephi}) respectively is non-axisymmetric, and therefore, it generates gravitation radiation as we will see below. In what follows, we will further assume that the source rotates slowly enough so that $r_0 \omega_0 \ll 1$. This is equivalent to say that the size $r_0$ of the source is much smaller than the light cylinder radius $R_{\rm lc}\equiv c/\omega_0$, or that the period of rotation $P$ is much larger than $2\pi r_0/c=0.2$~ms.

The Faraday tensor $F_{\mu\nu}$ can be constructed directly from
 %eqs.~(\ref{p0c}),
the expressions in eqs.~(\ref{Br})-(\ref{Ephi}),
%From eqs.~(\ref{p0b}) and eq.~(\ref{f2}) we obtain the components of the $F_{\mu\nu}$ tensor of the e/m
and using the equation~\cite{Thorne1982}
\begin{equation}\label{th1}
F^{ab}=u^a E^b-u^b E^a=\varepsilon^{abcd} u_c B_d
\end{equation}
where $u^a=(1,0,0,0)$ and $\varepsilon^{abcd}$ is the Levi-Civita symbol we obtain
\begin{equation}\label{f2}
F_{\mu\nu}=\left(
  \begin{array}{cccc}
    0 & -E^r & -r E^{\theta} & 0  \\
    E^r & 0 & r B^{\phi} & -r\sin{\theta} B^{\theta} \\
    r E^{\theta}&-r B^{\phi} & 0& r^2\sin{\theta}B^{r} \\
    0&r\sin{\theta} B^{\theta}& -r^2\sin{\theta}B^r &0
  \end{array}
\right)
\end{equation}
From eq.~(\ref{p0b}) and (\ref{f2}) we compute $T_{\mu\nu}$. We emphasize once again that we work in the slow rotation approximation where second and higher order powers of $r_0\omega_0$ are ignored. Therefore, according to eq.~(\ref{p0b}),
\begin{eqnarray}\label{t1}
T^{(1)}_{tt}&=&\frac{B_0^6 r_0^6}{8\pi r^6}(1+3\sin^2{\theta}\cos^2{w})\nonumber\\
T^{(1)}_{tr}&=&-\frac{B_0^2 r_0^6 \omega_0}{4\pi r^5} \sin{\theta}\sin{(2 w)}\nonumber\\
T^{(1)}_{t\theta}&=&\frac{B_0^2 r_0^6 \omega_0}{16\pi r^4} \sin{(2\theta)}\sin{(2w)}\nonumber\\
T^{(1)}_{t\phi}&=&-\frac{B_0^2 r_0^6 \omega_0}{8\pi r^4} \sin{(2\theta)}(1+3\sin^2{\theta}\cos^2{w})
\end{eqnarray}
\begin{eqnarray}\label{t2}
T^{(1)}_{rr}&=&\frac{B_0^2 r_0^6}{8\pi r^6}(1-5\sin^2{\theta}\cos^2{w})\nonumber\\
T^{(1)}_{r\theta}&=&\frac{B_0^2r_0^6}{4\pi r^5} \sin{(2\theta)}\cos^2{w}\nonumber\\
T^{(1)}_{r\phi}&=&\frac{B_0^2 r_0^6}{4\pi r^5} \sin^2{\theta}\sin{(2w)}
\end{eqnarray}
\begin{eqnarray}\label{t3}
T^{(1)}_{\theta\theta}&=&-\frac{B_0^2 r_0^6}{8\pi r^4} (\cos{(2w)}-5\sin^2{\theta}\cos^2{w})\nonumber\\
T^{(1)}_{\theta\phi}&=&\frac{B_0^2r_0^6}{16\pi r^4} \sin{(2\theta)}\cos{(2w)}\nonumber\\
T^{(1)}_{\phi\phi}&=&\frac{B_0^2r_0^6}{8\pi r^4} \sin^2{\theta}(\cos{(2w)}+3\sin^2{\theta}\cos^2{w})
\end{eqnarray}
where $w\equiv \omega_0 t-\phi$. The reader may check that the above energy momentum tensor has zero divergence in the limit $r_0^2\omega_0^2\simeq 0$ and zero trace , i.e. $Q^{\mu}\equiv T^{(1)}\ _{;\nu}^{\mu\nu}=0$ and $T^{(1)}\ _{\mu}^{\mu}=0$. Because of this result and since in flat space-time the d'Alembertian $\Box$ commutes with $\partial_{\mu}$, the Lorentz gauge condition reads
\begin{equation}\label{l1}
\partial^\nu \Box h_{\mu\nu}=\partial^{\nu} T^{(1)}_{\mu\nu}=0\Rightarrow \Box \partial^{\nu}h_{\mu\nu}=0\Rightarrow \partial^{\nu}h_{\mu\nu}=0
\end{equation}

Following the standard procedure appearing in the literature,~\cite{Wald1984,Maggiore2008,Bambi2022},
 the so-called TT-gauge may be applied, and the only independent components of the gravitational wave are $h_{\theta\theta}$, and $h_{\theta\phi}$. Thus, eqs.~(\ref{p0a}) yield two independent inhomogeneous wave equations
\begin{equation}\label{w1}
\Box h_{\theta\phi}=-16\pi T_{\theta\phi}^{(1)}
\end{equation}
\begin{equation}\label{w2}
\Box h_{\theta\theta}=-16\pi T_{\theta\theta}^{(1)}
\end{equation}
where $T^{(1)}_{\theta\theta}, T^{(1)}_{\theta\phi}$ are given by eqs.~(\ref{t3}). The source $T^{(1)}_{\phi\phi}$ yields one more component of the gravitational wave that depends on $h_{\phi\phi}$ and satisfies the inhomogeneous wave equation $\Box h_{\phi\phi}=-16\pi T_{\phi\phi}^{(1)}$, namely $h_{\phi\phi}=-\sin^2{\theta}\  h_{\theta\theta}$. We will next solve the above equations to determine $h_{\theta\phi}$ and $h_{\theta\theta}$.

\section{The Solution}

The inhomogeneous wave eqs.~(\ref{w1})-(\ref{w2}) are of the form
\begin{equation}\label{p1p}
\Box  \Psi =4\pi f\:,
\end{equation}
where $\Box $ is the d'Alembertian in any background, and $f\equiv f(t,r,\theta,\phi)$ is a source due to the electromagnetic field. In flat space time, eq.~(\ref{p1p}) reads
\begin{eqnarray}\label{p2}
&&\frac{\partial^2\Psi}{\partial t^2}-\frac{\partial^2 \Psi}{\partial r^2}-\frac{2}{r}\frac{\partial \Psi}{\partial r}\nonumber\\
&&-\frac{1}{r^2\sin{\theta}}\left(\frac{\partial}{\partial\theta}(\sin{\theta}\frac{\partial \Psi}{\partial\theta})+\frac{1}{\sin{\theta}}\frac{\partial^2 \Psi}{\partial\phi^2}\right)=4\pi f
\end{eqnarray}
Subsequently, we decompose $\Psi$ and the source term $f$ of eq.~(\ref{p2}) in spherical harmonics as
\begin{equation}\label{d2}
 \Psi =\sum_{lm}Y_{lm}(\theta,\phi)\int_{-\infty}^{\infty} d\omega e^{-i\omega t} R_{lm}(r_0,r;\omega)
\end{equation}
%and
\begin{equation}\label{d2f}
f=\sum_{lm}Y_{lm}(\theta,\phi)\int_{-\infty}^{\infty} d\omega e^{-i\omega t} f_{lm}(r_0,r;\omega)\ ,
\end{equation}
where
\begin{equation}\label{p21}
R_{lm}(r_{0},r ; \omega)= -i \omega H_l^{(1)}(\omega r) \int_{r_0}^{\infty}j_l(\omega \rho) f_{lm}(\rho;\omega)d\rho
\end{equation}
%and
\begin{equation}\label{p22p}
f_{lm}(r;\omega)=-2\int_{-\infty}^{\infty}dt e^{i\omega t}\int f(t,r,\theta,\phi)Y_{l m}^{*} (\theta,\phi)d\Omega\ .
\end{equation}
Here $H_l^{(1)}$  and $j_l$ are the Hankel and Bessel functions of the first kind respectively. Following the usual Fourier expansion method we end up with the inhomogeneous radial equation
\begin{equation}\label{p8}
\frac{d}{dr}\left( r^2\frac{dR_{lm}}{dr}\right)+(r^2 \omega^2 -l(l+1))R_{lm}=f_{lm}
%-2\int_{-\infty}^{\infty}g_{lm}(t,r)e^{i\omega t} dt
\end{equation}
A star on the spherical harmonics and other quantities on our equations means complex conjugate. Upon the consideration of eq.~(\ref{p21}), eq.~(\ref{d2}) becomes
\begin{equation}\label{d5}
\Psi(t,r,\theta,\phi)=-i\sum_{l,m} Y_{lm}(\theta,\phi)\int_{-\infty}^{\infty}\omega\ d\omega e^{-i\omega t}H_l^{(1)}(\omega r)\int_{r_{0}}^{\infty}j_l(\omega \rho) f_{lm}(\rho;\omega)d\rho
\end{equation}
Following the above decompositions, we are now ready to obtain the solutions of the inhomogeneous eqs.~(\ref{w1}) and (\ref{w2}).

%$$$$$$$$$$$$$$$$$$$$$$$$$$$$$$$$$$$$$$$$$$$$$$$$$$$$$$$$$$$$$$$$$$$$$$$$$$$$$$$$$$$$$$$$$$$$$$
\subsection{The solution of eq.~(\ref{w1})}
%$\Box h_{tt}=-2(B^r)^2[1+r^2\omega_0^2\sin^2{\theta}]$}
We start with eq.~(\ref{w1}), which in the limit $r_0^2\omega_0^2\simeq0$ and the aid of eqs.~(\ref{t3}) yields
\begin{equation}\label{k1p}
\Box h_{\theta\phi}=-\frac{B_0^2 r_0^6}{r^4}\sin{(2\theta)}\cos{(2w)},~~\mbox{~with~}~~w=\omega_0 t-\phi
\end{equation}
As it is apparent, its source term is
\begin{equation}\label{d6pr}
f=-\frac{B_0^2 r_0^6}{4\pi r^4}\sin{(2\theta)}\cos{(2w)}
\end{equation}
%%%%%%%%%%%%%%%%%%%%%%%%%%%%%%%%%%%%%%%%%%%%%%%%%%%%%%%%%%%%%%%%%%
%%%%%%%%%%%%%%%%%%%%%%%%%%%%%%%%%%%%%%%%%%%%%%%%%%%%%%
Eqs.~(\ref{p22p}) and (\ref{d6pr}) yield (see Appendix~C)
\begin{eqnarray}\label{d9x}
f_{lm}
&=&\frac{B_0^2r_0^6\sqrt{\frac{\pi}{5}}}{3r^4}\{\delta(\omega-2\omega_0)\int d\Omega Y_{l m }(\theta,\phi)\frac{d}{d\theta}(Y_{22}(\theta,\phi))\nonumber\\
&&+\ \delta(\omega+2\omega_0)\int d\Omega Y_{l m }(\theta,\phi)\frac{d}{d\theta}(Y_{22}^{*}(\theta,\phi))\}
\end{eqnarray}
%%%%%%%%%%%%%%%%%%%%%%%%%%%%%%%%%%%%%%%%%%%%%%%%%%%%%%%%%%%%%%%%%
%%%%%%%%%%%%%%%%%%%%%%%%%%%%%%%%%%%%%%%%%%%%%%%%%%%%%%%%%%

Furthermore, from eq.~(\ref{d5}) and eq.~(\ref{lat2}), we find the solution of eq.~(\ref{k1p})
%##################################
%######################################################
\begin{eqnarray}\label{solution2b}
&&h_{\theta\phi}(t,r,\theta,\phi;r_0,\omega_0)=i\frac{4\sqrt{5\pi}B_0^2 r_0^6}{3 r}\cdot\nonumber\\
&&\sum_{lm}\frac{2^l \omega_0^l}{(2l+1)!!}\frac{R_{\rm lc}^{(l-3)}-r_0^{(l-3)}}{l-3}Y_{lm}(\theta,\phi)S_{lm}\delta_{m2}e^{\frac{i}{2}[4t\omega_0-4\omega_0 r+l\pi]}\nonumber\\
\end{eqnarray}
where $S_{lm}$ is given by eq.~(\ref{w2p}) in Appendix~C.

We found that the double sum in eq.~(\ref{solution2b}) for $l=0,1$, and $m=-l$ to $m=+l\ $ becomes zero. In Appendix~C, we calculated the term for $l=2$  and found that it is also zero, since $S_{22}=0$. For  $l=3$ and $m=2$, we have $S_{32}=\frac{3}{128}\sqrt{210\pi}$, and using the de l'Hospital rule, we find that $(R_{\rm lc}^{(l-3)}-r_0^{(l-3)})/(l-3)$ approaches $\ln(r_0/R_{\rm lc})$ and thus
\begin{equation}\label{solutionc}
h_{\theta\phi}=3.1\times 10^{-3}\ \ln(\frac{r_0}{R_{\rm lc}})\frac{\sqrt{\pi}B_0^2 r_0^6\omega_0^3}{r}\cos(2[\omega_0 (t-r)+\phi])\cos\theta\sin^2\theta
\end{equation}
which is of order $(r_0\omega_0)^3$.

\subsection{The solution of eq.~(\ref{w2})}
%\subsection{The Solution of the Inhomogeneous Wave Equation:\\
%$\Box h_{rr}=-2(B^r)^2[1-r^2\omega_0^2\sin^2{\theta}]$}
This equation
\begin{equation}\label{s21}
\Box h_{\theta\theta}=-16\pi T_{\theta\theta}^{(1)}\equiv 4\pi f=\frac{2B_0^2 r_0^6}{r^4}(\cos{(2 w)}-5\sin^2{\theta}\cos^2{w})\ .
\end{equation}
Eq~(\ref{p22p}) yields
\begin{equation}\label{d7x}
f_{l m }=-\frac{B_0^2 r_0^6}{\pi r^4}\int_{-\infty}^{\infty} dt e^{i\omega t}\int(\cos{(2w)}-5\sin^2{\theta}\cos^2{w})
Y_{l  m }^{*}(\theta,\phi)d\Omega\nonumber\\
\end{equation}
%%%%%%%%%%%%%%%%%%%%%%%%%%%%%%%%%%%%%%%%%%%%%%
which becomes (see Appendix~D)
\begin{eqnarray}\label{qq5f}
f_{lm}(r,r_0;\omega;\omega_0)&=&-\frac{2\pi B_0^2 r_0^6}{r^4}\delta(\omega-2\omega_0)M_{lm}\delta_{m2}\nonumber\\
&+&\frac{5B_0^2 r_0^6}{r^4}\{\frac{1}{3}\sqrt{\frac{\pi}{5}}\delta(\omega-2\omega_0)\delta_{l2}\delta_{m2}\nonumber\\
&+&[\frac{4\sqrt{\pi}}{3}\delta_{l 0}\delta_{m0}-\frac{1}{\sqrt{5}}\delta_{l2}\delta_{m0}]\delta(\omega)\}
\end{eqnarray}
%################################
%In Appendix D, we calculate the
%\begin{equation}\label{g3b}
%I_{lm}=\int_{-\infty}^{\infty}\omega d\omega e^{-i\omega t}h_{l}^{(1)}(\omega r)\int_{r_{0}}^{R_{\rm lc}}J_{l}(\omega \rho) f_{lm}(\rho;\omega)d\rho
%\end{equation}
%needed to find the component $h_{\theta\theta}$. We found
%\begin{eqnarray}\label{g4x}
%I_{lm}(r,r_0;\omega,\omega_0)&=&\frac{iB_0^2 r_0^6}{6\pi r}\frac{2^l\omega_0^l}{(2l+1)!!}\frac{r_{0}^{l-3}-R_{\rm lc}^{l-3}}{l-3}\nonumber\\
%&\times&e^{i[2\omega_0 t-2\omega_0 r+\frac{l\pi}{2}]}[6\pi \delta_{m2} M_{lm}+%\sqrt{5\pi}\delta_{l2}\delta_{m2}]\nonumber\\
%\end{eqnarray}

%%%%%%%%%%%%%%%%%%%%%%%%%%%%%%%%%%%%%%%%%%%%%%%%%%
The solution of the inhomogeneous equation (\ref{s21}) now reads
\begin{eqnarray}\label{d5fa}
h_{\theta\theta}(t,r,\theta,\phi;r_0,\omega_0)&=&i\frac{B_0^2 r_0^6}{6\pi r}\sum_{lm}\frac{2^l \omega_0^l}{(2l+1)!!}\frac{R_{\rm lc}^{l-3}-r_0^{l-3}}{l-3}Y_{lm}(\theta,\phi)\nonumber\\
&&\cdot\  e^{i[2\omega_0 t-2\omega_0 r+\frac{l\pi}{2}]}[6\pi\delta_{m2}M_{lm}+\sqrt{5\pi}\delta_{l2}\delta_{m2}]\nonumber\\
\end{eqnarray}
where $M_{lm}$ is given in Appendix~D.
%\begin{equation}\label{af}
%M_{lm}=\sqrt{\frac{(2l+1)}{4\pi}\frac{(l-m)!}{(l+m)!}}\int_{-1}^{1} %P_l^m(\cos{\theta})d\cos{\theta}
%\end{equation}
Because of the form of eq.~(\ref{d5fa}), the first non-zero term of the double sum is obtained for $l=2$ and $m=2$. In this case $M_{22}=4$.
%$$$$$$$$$$$$$$$$$$$$$$$$$$$$$$$$$$$$$$$$$$$$$$$$
%##############################################

Inserting all values in eq.~(\ref{d5fa}) we find
\begin{eqnarray}\label{d5fd}
h_{\theta\theta}&=&\frac{B_0^2 r_0^6\omega_0^2\sin^2{\theta}}{15 r}\left(\frac{1}{R_{\rm lc}}-\frac{1}{r_0}\right)\sqrt{\frac{5}{\pi}}(24+\frac{5}{\sqrt{\pi}})\nonumber\\
&& \cdot\sin(2[\omega_0 (t-r)-\phi])
\label{htt}
\end{eqnarray}
and also
\begin{eqnarray}
h_{\phi\phi}&=&-\sin^2\theta\ h_{\theta\theta}
%\nonumber\\
%&=&\frac{B_0^2 r_0^6\omega_0^2\sin^4{\theta}}{15 r}\left(\frac{1}{R_{\rm lc}}-\frac{1}{r_0}\right)\sqrt{\frac{5}{\pi}}(24+\frac{5}{\sqrt{\pi}})\nonumber\\
%&& \cdot\sin(2[\omega_0 (t-r)-\phi])
\label{hpp}
\end{eqnarray}

\begin{figure}
\includegraphics[width=0.55\textwidth]{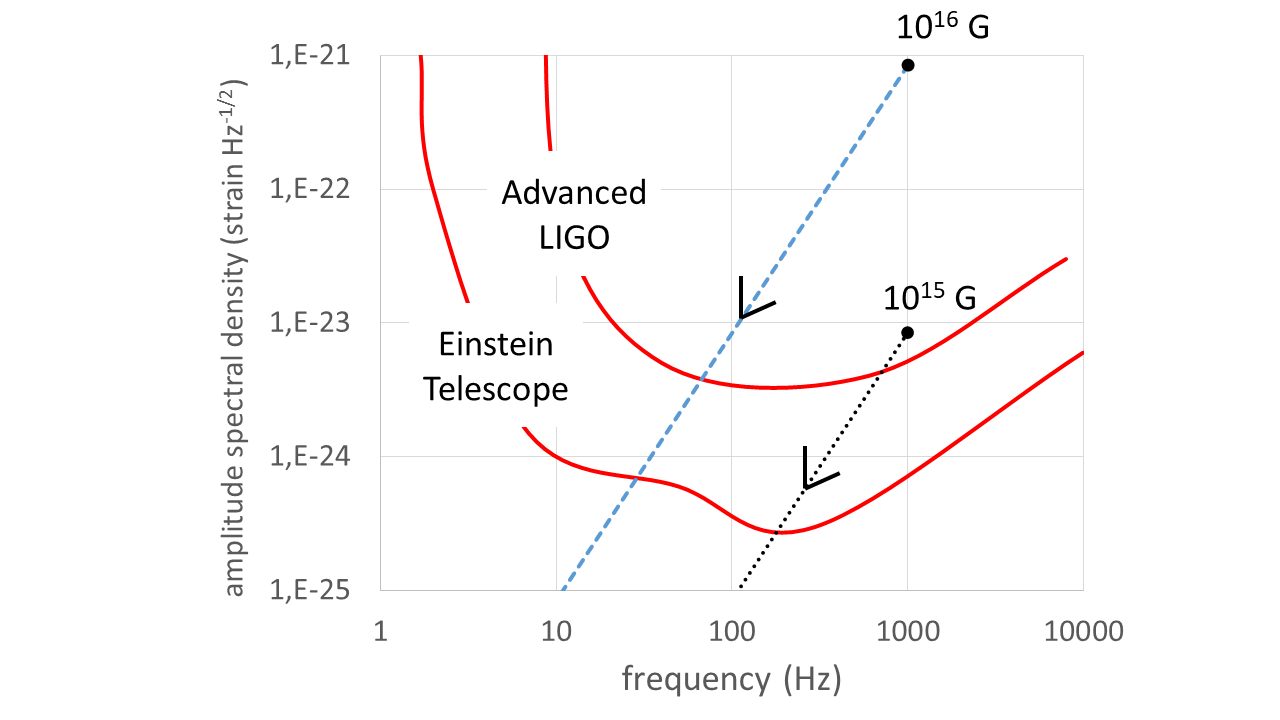}
    \caption{Thick red lines: approximate Advanced LIGO and Einstein Telescope sensitivity limits. Dashed line: amplitude of gravitational wave radiation on earth due to a $10^{16}$~G orthogonal magnetic rotator with initial rotation period of 1~ms (at neutron star formation) at a distance of 1~kpc. Dotted line: same for a $10^{15}$~G stellar magnetic field. The arrows show the direction of evolution of the signal as the pulsar spins down. The magnetic field contribution to the gravitational wave radiation of newly formed strongly magnetized pulsars is borderline observable with advanced gravitational wave detectors. }
 \label{Fig1}
 \end{figure}

\section{Discussion and Conclusions}
We found that, in the TT-gauge, the $h_{\theta\phi}$ and $h_{\theta\theta}$ components of the gravitational wave due to a rotating orthogonal dipole magnetic field behave as
\begin{equation}\label{r1}
h_{\theta\theta}\sim h_{\phi\phi}\sim (r_0  \omega_0)^2~~\mbox{~and~}~~h_{\theta\phi}\sim (r_0 \omega_0)^3
\end{equation}
Inserting real units in eqs.~(\ref{htt}) and (\ref{hpp}) we obtain
\begin{eqnarray}
h_{\theta\theta}&=&2.26\times\frac{G B_0^2 r_0^3}{c^4}\frac{1}{r}\frac{r_0^2 \omega_0^2 \sin^2{\theta}}{c^2}\sin(2[\omega_0 (t-r)-\phi])\nonumber\\
&=& 7\times 10^{-26}
\left(\frac{B_0}{10^{15}\ {\rm G}}\right)^2
\left(\frac{P}{10{\rm ms}}\right)^{-2}
\left(\frac{r}{{\rm kpc}}\right)^{-1}
\nonumber\\
&&\ \ \ \ \ \ \ \ \ \ \cdot \sin^2{\theta}\cdot\sin(2[\omega_0 (t-r)-\phi])
\label{result1}
\end{eqnarray}
and
\begin{eqnarray}
h_{\phi\phi}&=&-\sin^2\theta\ h_{\theta\theta}\nonumber\\
&=& -7\times 10^{-26}\
\left(\frac{B_0}{10^{15}\ {\rm G}}\right)^2
\left(\frac{P}{10{\rm ms}}\right)^{-2}
\left(\frac{r}{{\rm kpc}}\right)^{-1}
\nonumber\\
&&\ \ \ \ \ \ \ \ \ \ \cdot
\sin^4\theta\cdot\sin(2[\omega_0 (t-r)-\phi])
\label{result2}
\end{eqnarray}
For galactic pulsars that are born with periods $P\sim 1\ {\rm ms}$, radius $r_0=12$~km and extra strong magnetic fields $B_0> 10^{15}\ {\rm G}$ within a distance of about 1~kpc, this gravitational signal is borderline observable by the new generation of gravitational wave detectors (Advanced LIGO and Einstein Telescope; see figure~\ref{Fig1})
around the time of the neutron star formation when their frequency of rotation is higher than about 100~Hz. 
If observed, the shape of the gravitational wave signal would yield details about the distribution of the magnetic field in the inner neutron star magnetosphere. We have assumed here an orthogonal dipole field magnetosphere, but our results may be generalized for higher order multipole magnetic field configurations (quadrupole, etc.). It may be difficult to differentiate between gravitational wave signals coming from the magnetosphere and from a potential non-sphericity of the neutron star matter distribution. Notice that a result similar to eqs.~(\ref{result1}) and (\ref{result2}) was obtained in \cite{Contopoulos2022} with an order of magnitude calculation of the magnetic field asymmetry in the gravitational collapse of supermassive star cores. In that case, we predicted a particular waveform for the resulting gravitation wave (figure~5 in that paper). In the present case of a young strongly magnetized neutron star, the waveform will possess consistent frequency and amplitude, akin to a singer holding a sustained note.

{
As we acknowledged in the Introduction, Nazari \& Roshan~(2020) solved essentially the same problem, namely the direct contribution of a rotating dipolar magnetic field to the generation of gravitational waves. Their results are more general than ours since they consider all inclinations of the dipole magnetic field with respect to the axis of rotation, but their method is different. They implement a PN approximation to the energy–momentum tensor of the magnetized fluid and to the gravitational potential in the wave zone. We instead work directly with the perturbation of the stress-energy tensor $T^{(1)}_{\mu\nu}$ (eq.~2), and study its effect on the background metric $h_{\mu\nu}$ (eq.~1) by solving the wave equation that connects the two (eq.~14) in the TT-gauge. In principle, the two methods are expected to yield the same result in the far zone. However, our final estimate of the strain of the gravitational wave observed on earth (eqs.~\ref{result1} \& \ref{result2}) is two orders of magnitude higher than their corresponding numerical estimate (their eq.~48). We conclude that the gravitational wave due to the rotating electromagnetic field is borderline observable with the new generation of gravitational wave detectors, while they conclude that it is unobservable and in order to obtain an observable signal they resort to signal-to-noise (SN) amplification with integration times of several days (Moore, Cole \& Berry 2014).

Their main result in their eq.~(47) looks very similar to ours in eqs.~(\ref{result1}) and (\ref{result2}), only the orders of magnitude differ. After checking their calculations more carefully we realized that their magnetic field value is the polar value of the stellar dipole magnetic field which is double our $B_0$ value. When we recalculated ourselves their result we obtained a different numerical factor, namely
\begin{eqnarray}
h_{0F} & = & 
\frac{32}{105}\frac{\pi^3 G}{\mu_0 c^6}\frac{(2B_0)^2 r_0^5}{P^2 r}
\nonumber\\
&=& 3\times 10^{-26}\
\left(\frac{B_0}{10^{15}\ {\rm G}}\right)^2
\left(\frac{P}{10{\rm ms}}\right)^{-2}
\left(\frac{r}{{\rm kpc}}\right)^{-1}
\label{h0F}
\end{eqnarray}
which is now of the same order of magnitude as our result (notice that the magnetic permeability of vacuum $\mu_0=1$ in the CGS units that we are both using). These extra two orders of magnitude that we obtained in this recalculation allow us to claim that our results are in agreement with the corrected estimates of Nazari \& Roshan~(2020).

We decided to further check the order of magnitude of our result through an analogy with the calculation of the strain of the gravitational wave $h_M$ generated by a rotating deformed neutron star. In that case,
\begin{eqnarray}
h_M&=&\frac{4G I_{zzM}\Omega^2}{c^4r}\epsilon_M=
\frac{8GM_* r_0^2 \Omega^2}{5c^4 r}\epsilon_M\nonumber\\
&=& 7\times 10^{-20}\epsilon_M
\left(\frac{P}{10{\rm ms}}\right)^{-2}
\left(\frac{r}{{\rm kpc}}\right)^{-1}
\label{hM}
\end{eqnarray}
(eq.~ 31a from Nazari \& Roshan~2020), where, 
\begin{eqnarray}
I_{zzM} &=& \int_{r=0}^{r_0} \int_{\theta=0}^\pi \int_{\phi=0}^{2\pi}
r^2 \sin\theta\ dr\ d\theta\ d\phi\ \rho_M(x^2+y^2)\approx
\frac{2}{5}M_* r_0^2\\
\label{eM}
I_{xxM} &=& \int_{r=0}^{r_0} \int_{\theta=0}^\pi \int_{\phi=0}^{2\pi}
r^2 \sin\theta\ dr\ d\theta\ d\phi\ \rho_M(y^2+z^2)\\
\epsilon_M & = & 1-\frac{I_{xxM}}{I_{zzM}}
\label{Izz}
\end{eqnarray}
Here, $I_{zzM}$ is the stellar matter moment of inertia around the $z$ axis of rotation, $\epsilon_M$ is the ellipticity of the stellar matter distribution (eq.~4 of \cite{Mastranoetal2015}), and 
$M_*\approx 1.4M_\odot$ is the mass of the neutron star. The main difference between our calculation and eqs.~(\ref{hM}-\ref{Izz}) is that, in our case, the source of the gravitational wave is the energy density distribution {\it outside} the surface of the star, therefore, we must calculate the above volume integrals for $r>r_0$. If we replace $\rho_M$ in eqs.~(\ref{eM}-\ref{Izz}) by $B^2/(8\pi c^2)=[(B^r)^2+(B^\theta)^2+(B^\phi)^2]/(8\pi c^2)$ where the magnetic field components are given by eqs.~(\ref{Br}-\ref{Bphi}) for $t=0$, and we also replace the radial limits of integration from $(0,r_0)$ to $(r_0,\infty)$ we obtain
\begin{eqnarray}
I_{zzF} &=& \int_{r=r_0}^{\infty} \int_{\theta=0}^\pi \int_{\phi=0}^{2\pi}
r^2 \sin\theta\ dr\ d\theta\ d\phi\ B^2(x^2+y^2)/(8\pi c^2)\nonumber\\
& = & \frac{11}{15}\frac{B_0^2 r_0^5}{c^2}
\label{eF}\\
I_{xxF} &=& \int_{r=r_0}^{\infty} \int_{\theta=0}^\pi \int_{\phi=0}^{2\pi}
r^2 \sin\theta\ dr\ d\theta\ d\phi\ B^2(y^2+z^2)/(8\pi c^2)\nonumber\\
& = & \frac{8}{15}\frac{B_0^2 r_0^5}{c^2}\\
\epsilon_F & = & 1-\frac{I_{xxF}}{I_{zzF}}=\frac{3}{11}=0.27\ .
\label{IzzF}
\end{eqnarray}
Using the above expressions, we can obtain an order of magnitude estimate of the electromagnetic contribution to the gravitational wave signal  $h_F$ through an analogy with the calculation that led to eq.~(\ref{hM}), namely
\begin{eqnarray}
h_F&\sim & h_M\frac{I_{zzF}}{I_{zzM}}\frac{\epsilon_{F}}{\epsilon_{M}}= \frac{1}{2}
\left(\frac{B_0^2 r_0^3}{M_* c^2}\right)
\frac{8GM_* r_0^2 \Omega^2}{5c^4 r}\nonumber\\
& = &
2.4\times 10^{-26}\
\left(\frac{B_0}{10^{15}\ {\rm G}}\right)^2
\left(\frac{P}{10{\rm ms}}\right)^{-2}
\left(\frac{r}{{\rm kpc}}\right)^{-1}\ .
\end{eqnarray}
This estimate is of the same order of magnitude as our results in eqs.~(\ref{result1}), (\ref{result2}) and (\ref{h0F}).
}

In the present work we considered the perturbation of the electromagnetic field energy on the Minkowski (empty) space time. In a forthcoming publication we will investigate the generation of gravitational waves from rotating neutron star magnetospheres in the Schwarzschild and the Kerr space time.

\section*{Data availability statement}
The data underlying this article will be shared on reasonable request to the corresponding author.

%$$$$$$$$$$$$$$$$$$$$$$$$$$$$$$$$$$$$$$$$$$$$$
%\newpage\
{}

\section*{Appendix A}

We will be searching for the steady-state solution to eqs.~(\ref{em2a}) in spherical coordinates $(r,\theta,\phi)$ centered onto the central star. We will follow the approach of \cite{Muslimov2009}, and we will define the transformation of partial time derivatives between our lab frame, and a reference frame that rotates with the pulsar angular velocity $\omega_0$ around the axis $\theta=0$, namely
\begin{eqnarray}
\frac{\partial {\bf B}}{\partial t} & = &
\left.\frac{\partial {\bf B}}{\partial t}\right|_{\rm corot}-
\nabla\times (\omega_0 r\sin\theta\ \hat{\phi}\times {\bf B})\ ,\\
\frac{\partial {\bf E}}{\partial t} & = &
\left.\frac{\partial {\bf E}}{\partial t}\right|_{\rm corot}-
\nabla\times (\omega_0 r\sin\theta\ \hat{\phi}\times {\bf E})
+\omega_0 r\sin\theta\ \hat{\phi}\ \nabla\cdot{\bf E}\ ,\\
\frac{\partial \rho_{\rm e}}{\partial t} & = &
\left.\frac{\partial \rho_{\rm e}}{\partial t}\right|_{\rm corot}
+\omega_0 r\sin\theta\ \hat{\phi}\cdot\nabla\rho_{\rm e}\ .
\end{eqnarray}
Time derivatives in that frame (denoted by the subscript ``corot'') vanish in steady state, and therefore, Maxwell's equations that describe the steady state of the force-free pulsar magnetosphere become
\begin{eqnarray}
\nabla\times\left({\bf E}+\frac{\omega_0 r\sin\theta\ \hat{\phi}}{c}\times{\bf B}\right) &=& 0\ ,\label{E}\\
\nabla\times\left({\bf B}-\frac{\omega_0 r\sin\theta\ \hat{\phi}}{c}\times{\bf E}\right) &=& \frac{4\pi}{c}{\bf J}-\frac{\omega_0 r\sin\theta\ \hat{\phi}}{c}\nabla\cdot{\bf E}\ .\label{B}
\end{eqnarray}
From eq.~(\ref{E}) above we obtain that
\begin{eqnarray}
&&{\bf E}=-\frac{\omega_0 r\sin\theta}{c}\hat{\phi}\times{\bf B}\ .
\label{E2}
\end{eqnarray}

\section*{Appendix B: Usefull relations}
We use the following relations from \cite{Arfken2001} for Legendre polynomials $P_n(x)$ and spherical harmonics $Y_{lm}(\theta,\phi)$:
\begin{equation}\label{I1}
P_n(x)=\frac{1}{2^n n!}(\frac{d}{dx})^n (x^2-1)^n
\end{equation}
\begin{equation}\label{l2}
P_n^m(x)=(1-x^2)^{m/2}(\frac{d}{dx})^m P_n(x)
\end{equation}
\begin{equation}\label{a0}
Y_{lm}(\theta,\phi)=\sqrt{\frac{2l+1}{4\pi}\frac{(l-m)!}{(l+m)!}}P_l^{m}(\cos{\theta})e^{im\phi}
\end{equation}
and $Y_{l,-m}=(-1)^m Y^{*}_{lm}$ where $Y^{*}_{lm}$ is the complex conjugate of $Y^{*}_{lm}$.
From eq.~(\ref{a0}) we compute the
\begin{eqnarray}\label{a1}
Y_{00}(\theta,\phi)&=&\frac{1}{\sqrt{4\pi}}\nonumber\\
Y_{20}(\theta,\phi)&=&\frac{1}{2}\sqrt{\frac{5}{\pi}}P_2(\cos{\theta})\nonumber\\
%Y_{40}(\theta,\phi)&=&\frac{3}{2}\sqrt{\frac{1}{\pi}}P_4(\cos{\theta})=\frac{3}{2}\sqrt{\frac{1}{\pi}}[\frac{35}{8}\sin^4{\theta}-5\sin^2{\theta}+1]\nonumber\\
%Y_{60}(\theta,\phi)&=&\frac{1}{2}\sqrt{\frac{13}{\pi}}P_6(\cos{\theta})=\frac{1}{2}\sqrt{\frac{13}{\pi}}[-\frac{231}{16}\sin^6{\theta}+\frac{189}{8}\sin^4{\theta}-\frac{21}{2}\sin^2{\theta}+1]\nonumber\\
Y_{22}(\theta,\phi)&=&\frac{1}{2}\sqrt{\frac{5}{\pi}}P_2^2(\cos{\theta})e^{2i\phi}=\frac{3}{2}\sqrt{\frac{5}{\pi}}\sin^2{\theta}e^{2i\phi}\nonumber\\
%Y_{42}(\theta,\phi)&=&\frac{1}{40}\sqrt{\frac{18}{\pi}}P_{4}^{2}(\cos{\theta})e^{2i\phi}=\frac{1}{40}\sqrt{\frac{18}{\pi}}e^{2i\phi}[-\frac{105}{2}\sin^4{\theta}+45\sin^2{\theta}]\nonumber\\
%Y_{62}(\theta,\phi)&=&\frac{1}{840}\sqrt{\frac{1365}{\pi}}P_6^{2}(\cos{\theta})e^{2i\phi}=\frac{1}{840}\sqrt{\frac{1365}{\pi}}e^{2i\phi}[\frac{3465}{8}\sin^6{\theta}+630\sin^4{\theta}+216\sin^2{\theta}]\nonumber\\
\end{eqnarray}
\begin{eqnarray}\label{a1x}
\sin^2{\theta}&=&\frac{4\sqrt{\pi}}{3}[Y_{00}(\theta,\phi)-\frac{Y_{20}(\theta,\phi)}{\sqrt{5}}]\nonumber\\
%\sin^4{\theta}&=&\frac{16\sqrt{\pi}}{105}Y_{40}(\theta,\phi)-\frac{160}{105}\sqrt{\frac{\pi}{5}}Y_{20}(\theta,\phi)-\frac{112\sqrt{\pi}}{105} Y_{00}(\theta,\phi)\nonumber\\
%\sin^6{\theta}&=&-\frac{32}{231}\sqrt{\frac{\pi}{13}}Y_{60}(\theta,\phi)+\frac{96\sqrt{\pi}}{385}Y_{40}(\theta,\phi)-\frac{32}{31}\sqrt{\frac{\pi}{5}}Y_{20}(\theta,\phi)-\frac{992\sqrt{\pi}}{385}Y_{00}(\theta,\phi)\nonumber\\
%\end{eqnarray}
%\begin{eqnarray}\label{a2s}
\sin^2{\theta}e^{2i\phi}&=&\frac{2}{3}\sqrt{\frac{\pi}{5}}Y_{22}(\theta,\phi)\nonumber\\
%\sin^4{\theta}e^{2i\phi}&=&-\frac{80}{105}\sqrt{\frac{\pi}{10}} Y_{42}(\theta,\phi)                               +\frac{60}{105}\sqrt{\frac{\pi}{5}}Y_{22}(\theta,\phi)\nonumber\\
%\sin^6{\theta}e^{2i\phi}&=&\frac{64}{33}\sqrt{\frac{\pi}{1365}}Y_{62}(\theta,\phi)-\frac{256}{231}\sqrt{\frac{\pi}{10}}Y_{42}(\theta,\phi)+\frac{32}{63}\sqrt{\frac{\pi}{5}}Y_{22}(\theta,\phi)
\end{eqnarray}
\newline\
%%%%%%%%%%%%%%%%%%%%%%%%%%%%%%%%%%%%%%%%%%%%%%%%%%%%%%%%%%%%%%%%%%%%%%%%%%%%%%%%%%%%

In our calculations we also use the spherical Bessel functions $j_l(x), n_l(x)$ and the Hankel functions $H_l^{(1)}, H_l^{(2)}$ defined as
%either in their exact form or in their approximate one.
\begin{equation}\label{ap1}
H_l^{(1)}(x)\equiv j_l(x)+i n_l(x),~~H_l^{(2)}(x)\equiv j_l(x)-i n_l(x)
\end{equation}

%$$$$$$$$$$$$$$$$$$$$$$$$$$$$$$$$$$$$$$$$$$$
For approximate calculations, we may use limiting values of the above fucntions to calculate the integrals in eq.~(\ref{d5}). For $x\rightarrow\infty$, we have
\begin{eqnarray}\label{p12pw}
&&j_l(x)=\frac{1}{x}\sin\left(x-\frac{l \pi}{2}\right)\nonumber\\
%n_{l}(\omega r)&=&-\sin{[\omega r-\frac{\pi}{2}(l+1)+\frac{l(l+1)}{2\omega r}]}+O(1/\omega^2 r^2
&&H_l^{(1)}(x)=-\frac{i}{x} \exp\left(-i(x-\frac{l \pi}{2})\right)\nonumber\\
\end{eqnarray}
Also, for $x<< 1$, we have
\begin{eqnarray}\label{p13}
&&j_l(x)\approx\frac{x^l}{(2l+1)!!}\nonumber\\
&&\eta_l(x)\approx-\frac{((2l-1)!!}{x^{l+1}}\nonumber\\
&&H_{l}^{(1),(2)}\approx \pm i\eta_l(x)
\end{eqnarray}
\cite{Arfken2001}. In our further calculations we use eqs.~(\ref{p13}) in the limit $x=\omega_0 r\ll 1$.

\section*{Appendix C}
eqs.~(\ref{p22p}) and (\ref{d6pr}) yield
\begin{eqnarray}\label{d7kk}
f_{lm}(r,r_0;\omega;\omega_0)&=&-2\int_{-\infty}^{\infty} dt e^{i\omega t}\int(-\frac{B_0^2 r_0^6}{4\pi r^4}\sin{(2\theta)}\cos{(2w)} ) Y_{l m}^{*}(\theta,\phi)d\Omega\nonumber\\
&=&\frac{B_0^2 r_0^6}{2\pi r^4}\int_{-\infty}^{\infty} dt e^{i\omega t}\int\sin{(2\theta)}[\frac{e^{2iw}+e^{-2iw}}{2}]Y_{l m}^{*}(\theta,\phi)d\Omega
\nonumber\\
%&=&\frac{B_0^2 r_0^6}{4\pi r^4}\int_{-\infty}^{\infty} dt e^{i\omega t}\int\sin{2\theta}[e^{-2i\omega_0 t}e^{2i\phi}+e^{2i\omega_0 t}e^{-2i\phi}]Y_{l m}^{*}(\theta,\phi)d\Omega\nonumber\\
\end{eqnarray}
We define
\begin{eqnarray}\label{d10a}
W_{lm}^{(1)}&\equiv&\frac{B_0^2 r_0^6}{4\pi r^4}\int_{-\infty}^{\infty} dt e^{it(\omega-2\omega_0)}\int\sin{(2\theta)}e^{2i\phi}Y_{lm}^{*}(\theta,\phi) d\Omega\nonumber\\
&=&\frac{B_0^2 r_0^6}{2 r^4})\delta(\omega-2\omega_0)\int\sin{(2\theta)}e^{2i\phi}Y_{lm}^{*}(\theta,\phi) d\Omega
\end{eqnarray}
and
\begin{eqnarray}\label{d10b}
W_{lm}^{(2)}&\equiv&\frac{B_0^2 r_0^6}{4\pi r^4}\int_{-\infty}^{\infty} dt e^{it(\omega+2\omega_0)}\int\sin{(2\theta)}e^{-2i\phi}Y_{lm}^{*}(\theta,\phi) d\Omega\nonumber\\
&=&\frac{B_0^2 r_0^6}{2 r^4}\delta(\omega+2\omega_0)\int\sin{(2\theta)}e^{-2i\phi}Y_{lm}^{*}(\theta,\phi) d\Omega
\end{eqnarray}

Using the relations
\begin{eqnarray}\label{d11z}
&&2\sin{\theta}cos{\theta}e^{2i\phi}=\frac{2}{3}\sqrt{\frac{\pi}{5}}\frac{d}{d\theta}Y_{22}(\theta,\phi)\ ,\nonumber\\
&&2\sin{\theta}cos{\theta}e^{-2i\phi}=\frac{2}{3}\sqrt{\frac{\pi}{5}}\frac{d}{d\theta}Y_{22}^{*}(\theta,\phi)\ .\nonumber\\
\end{eqnarray}
eq.~(\ref{d7kk})  becomes
\begin{eqnarray}\label{d7km}
f_{lm}(r,r_0;\omega;\omega_0)&=&W_{lm}^{(1)}+W_{lm}^{(2)}\nonumber\\
&=&\frac{B_0^2 r_0^6}{2 r^4}\{\delta(\omega-2\omega_0)\int\sin{(2\theta)}e^{2i\phi}Y_{lm}^{*}(\theta,\phi) d\Omega\nonumber\\
&&+\delta(\omega+2\omega_0)\int\sin{(2\theta)}e^{-2i\phi}Y_{lm}^{*}(\theta,\phi) d\Omega\}\nonumber\\
&=&\frac{B_0^2 r_0^6}{3 r^4}\sqrt{\frac{\pi}{5}}\nonumber\\
&& \cdot\ \{\delta(\omega-2\omega_0)\int d\Omega Y_{lm}^{*}(\theta,\phi)\frac{d}{d\theta}Y_{22}(\theta,\phi)\nonumber\\
&&+\delta(\omega+2\omega_0)\int d\Omega Y_{lm}^{*}(\theta,\phi)\frac{d}{d\theta}Y_{22}^{*}(\theta,\phi)\}\nonumber\\
\end{eqnarray}
We define
\begin{equation}\label{u3}
A_{lm}\equiv\int d\Omega Y_{lm}^{*}(\theta,\phi)\frac{d}{d\theta}Y_{22}(\theta,\phi)
\end{equation}
and
\begin{equation}\label{u4}
\tilde{A}_{lm}\equiv \int d\Omega Y_{lm}^{*}(\theta,\phi)\frac{d}{d\theta}Y_{22}^{*}(\theta,\phi)
\end{equation}
To  calculate the $A_{lm}$ and $\tilde{A}_{lm}$, we use the orthogonality condition
\begin{equation}\label{arf1}
\int_0^{2\pi}e^{-im_1\phi} e^{im_2\phi} d\phi=2\pi\delta_{m_1 m_2}
\end{equation}
\cite{Arfken2001} and  we find
\begin{eqnarray}\label{w1p}
A_{lm}&=&2\int\sin{\theta}\cos{\theta} e^{2i\phi} Y_{lm}^{*}(\theta,\phi) d\Omega\nonumber\\
&=&2\sqrt{\frac{(2l+1)}{4\pi}\frac{(l-m)!}{(l+m)!}}\nonumber\\
&&\cdot\ \int_0^{2\pi}e^{2i\phi} e^{-im\phi}d\phi \int_0^{\pi}\sin{\theta}\cos{\theta} P_{l}^{m}(\cos{\theta})\sin{\theta}d\theta\nonumber\\
&=&4\pi\delta_{m2}\sqrt{\frac{(2l+1)}{4\pi}\frac{(l-m)!}{(l+m)!}}\nonumber\\
&&\cdot\ \int_{-1}^{1} \frac{1}{3}P_2^1(\cos{\theta})P_{l}^{m}(\cos{\theta})d\cos{\theta}
\end{eqnarray}
For simplicity we set
\begin{eqnarray}\label{w2p}
S_{lm}&\equiv&\sqrt{\frac{(2l+1)}{4\pi}\frac{(l-m)!}{(l+m)!}}\int_{-1}^{1} \frac{1}{3}P_2^1(\cos{\theta})P_{l}^{m}(\cos{\theta})d\cos{\theta}\nonumber\\
&=&\sqrt{\frac{(2l+1)}{4\pi}\frac{(l-m)!}{(l+m)!}}\int_{-1}^{1} \cos{\theta}\sqrt{1-\cos^2{\theta}}P_{l}^{m}(\cos{\theta})d\cos{\theta}\nonumber\\
\end{eqnarray}
and eq.~(\ref{w1p}) reads
\begin{equation}\label{w3p}
A_{lm}=4\pi\delta_{m2}S_{lm}
\end{equation}
%###############################################
%$$$$$$$$$$$$$$$$$$$$$$$$$$$$$$$$$$$$$$$$$$$$$
Because of the relation
\begin{equation}\label{arf2p}
\int_0^{2\pi}e^{-im_1\phi} e^{-im_2\phi} d\phi=0
\end{equation}
$\tilde{A}_{lm}=0$. Eventually,   eq.~(\ref{d7km}) becomes
\begin{eqnarray}\label{d7kf}
f_{lm}(r,r_0;\omega;\omega_0)&=&\frac{B_0^2 r_0^6}{2 r^4}\delta(\omega-2\omega_0)\int\sin{(2\theta)}e^{2i\phi}Y_{lm}^{*}(\theta,\phi) d\Omega\nonumber\\
&=&\frac{B_0^2 r_0^6}{3 r^4}\sqrt{\frac{\pi}{5}}\delta(\omega-2\omega_0)4\pi\delta_{m2}S_{lm}
\end{eqnarray}
%$$$$$$$$$$$$$$$$$$$$$$$$$$$$$$$$$$$$$$$$$$$$$$$$$$$$$$$$$
%%%%%%%%%%%%%%%%%%%%%%%%%%%%%%%%%%%%%%%%
As we mention in Sec.3,  the solution of the inhomogeneous eq.~(\ref{w1}) will be of the form of eq.~(\ref{d5}). Therfore, we need to compute the following integrals:
%Using eq.~(\ref{p12pw}), the integral in eq.~(\ref{d7kf}) becomes
\begin{eqnarray}\label{ju2}
J_{lm}(r_0;\omega;\omega_0)&=&\int_{r_0}^{R_{\rm lc}} j_{l}(\omega r) f_{lm}(r;\omega;\omega_0)dr\nonumber\\
&=&\frac{B_0^2 r_0^6}{3}\sqrt{\frac{\pi}{5}}\int_{r_0}^{R_{\rm lc}}\frac{\omega^l r^l}{r^4(2l+1)!!}[A_{lm}\delta(\omega-2\omega_0)]dr\nonumber\\
&=&-\frac{B_0^2 r_0^6}{3}\sqrt{\frac{\pi}{5}}\frac{\omega^l }{(2l+1)!!}\frac{r_0^{(l-3)}-R_{\rm lc}^{(l-3)}}{l-3}4\pi\delta_{lm}S_{lm}\delta(\omega-2\omega_0)\nonumber\\
\end{eqnarray}
and
%Using eqs.~(\ref{p13}) and (\ref{u2}) we also found that
\begin{eqnarray}\label{lat2}
I_{lm}&=&\int_{-\infty}^{\infty}\omega d\omega e^{-i\omega t} H_{l}^{(1)}(\omega r) J_{lm}(r_0;\omega,\omega_0)\nonumber\\
&=&-\frac{B_0^2 r_0^6}{3 r}\sqrt{\frac{5}{\pi}}\frac{2^l \omega_0^l}{(2l+1)!!}\frac{r_{0}^{(l-3)}-R_{\rm lc}^{(l-3)}}{l-3}\nonumber\\
&&\cdot\  4\pi\delta_{lm}S_{lm}e^{\frac{i}{2}[4t\omega_0-4\omega_0 r+l\pi]}
\end{eqnarray}

Now, as it is apparent,  with the aid of eqs.~(\ref{ju2})  and (\ref{lat2})  we can find the solution of the inhomogeneous eq.~(\ref{k1p}).
%%%%%%%%%%%%%%%%%%%%%%%%%%%%%%%%%%%%%%%%%%%%%%%%%%%%%%%%%%%%%

\section*{Appendix D}
We re-write
\begin{eqnarray}
f_{lm}(r,r_0;\omega;\omega_0)&=&-2\int_{-\infty}^{\infty} dt e^{i\omega t}\nonumber\\
&&\cdot\ \int\frac{B_0^2 r_0^6}{2\pi r^4}[\cos{(2w)}-5\sin^2{\theta}\cos^2{w}]
Y_{l m}^{*}(\theta,\phi)d\Omega\nonumber\\
\label{d7xx}
\end{eqnarray}
%%%%%%%%%%%%%%%%%%%%%%%%%%%%%%%%%%%%%%
%$$$$$$$$$$$$$$$$$$$$$$$$$$$$$$$$$$$$$$$$$$$$$
We define
\begin{eqnarray}\label{q1}
Q_{lm}^{(1)}&\equiv&  -\frac{B_0^2 r_0^6}{\pi r^4}\int_{-\infty}^{\infty} dt e^{i\omega t}\int[\frac{e^{2iw}+e^{-2iw}}{2}]Y_{l m}^{*}(\theta,\phi)d\Omega\nonumber\\
Q_{lm}^{(2)}&\equiv & \frac{5B_0^2 r_0^6}{\pi r^4}\int_{-\infty}^{\infty} dt e^{i\omega t}\int \sin^2{\theta}\cos^2{w}
Y_{l m}^{*}(\theta,\phi)d\Omega
\end{eqnarray}
Below, we calculate $Q_{lm}^{(1)}$ and $Q_{lm}^{(2)}$ and upon the consideration of  eqs.~(\ref{a0}), (\ref{arf1}) and (\ref{arf2p}), we find
\begin{eqnarray}\label{qq1}
Q_{lm}^{(1)}&\equiv&  -\frac{B_0^2 r_0^6}{\pi r^4}\int_{-\infty}^{\infty} dt e^{i\omega t}\int[\frac{e^{2iw}+e^{-2iw}}{2}]Y_{l m}^{*}(\theta,\phi)d\Omega\nonumber\\
&=&-\frac{B_0^2 r_0^6}{2\pi r^4}\int_{-\infty}^{\infty} dt e^{i\omega t}\int[e^{-2i\omega_0 t} e^{2i\phi}+e^{2i\omega_0 t} e^{-2i\phi}]Y_{l m}^{*}(\theta,\phi)d\Omega\nonumber\\\nonumber\\
&=&-\frac{B_0^2 r_0^6}{2\pi r^4}\{2\pi\delta(\omega-2\omega_0)\int e^{2i\phi}Y_{l m}^{*}(\theta,\phi)d\Omega\nonumber\\
&& +2\pi\delta(\omega+2\omega_0)\int e^{-2i\phi}Y_{l m}^{*}(\theta,\phi)d\Omega\}\nonumber\\&=&  -\frac{B_0^2 r_0^6}{r^4}\delta(\omega-2\omega_0)\int e^{2i\phi}Y_{l m}^{*}(\theta,\phi)d\Omega\nonumber\\
\end{eqnarray}
%%%%%%%%%%%%%%%%%%%%%%%%%%%%%%%%%%%%%%%
%%%%%%%%%%%%%%%%%%%%%%%%%%%%%%%%%%%%%%%%%%%
Working as previously and taking into account eqs.~(\ref{a0}), (\ref{arf1}) and (\ref{arf2p}) we find
\begin{eqnarray}\label{qq2x}
Q_{lm}^{(2)}&\equiv & \frac{5B_0^2 r_0^6}{\pi r^4}\int_{-\infty}^{\infty} dt e^{i\omega t}\int[\sin^2{\theta}\cos^2{w}]
Y_{l m}^{*}(\theta,\phi)d\Omega\nonumber\\
&=&\frac{5B_0^2 r_0^6}{\pi r^4}\int_{-\infty}^{\infty} dt e^{i\omega t}\int\sin^2{\theta}Y_{l m}^{*}(\theta,\phi)\frac{e^{2iw}+e^{-2iw}+2}{4}\nonumber\\
&=&\frac{5B_0^2 r_0^6}{4\pi r^4}\int_{-\infty}^{\infty} dt e^{i\omega t}
\int\sin^2{\theta}Y_{l m}^{*}(\theta,\phi)[e^{-2i\omega_0 t} e^{2i\phi}+e^{2i\omega_0 t}e^{-2i \phi}+2]\nonumber\\
&=&\frac{5B_0^2 r_0^6}{4\pi r^4}\{2\pi\delta(\omega-2\omega_0)\frac{2}{3}\sqrt{\frac{\pi}{5}}\int d\Omega Y_{lm}^{*}(\theta,\phi)Y_{22}(\theta,\phi)\nonumber\\
&&+2\pi\delta(\omega+2\omega_0)\frac{2}{3}\sqrt{\frac{\pi}{5}}\int d\Omega Y_{lm}^{*}(\theta,\phi)Y_{22}^{*}(\theta,\phi)\nonumber\\
&&+4\pi\delta(\omega)\int d\Omega Y_{lm}^{*}(\theta,\phi)\sin^2{\theta}\}\nonumber\\&=& \frac{5B_0^2 r_0^6}{2 r^4}\{\frac{2}{3}\sqrt{\frac{\pi}{5}}\delta(\omega-2\omega_0)\delta_{l 2}\delta_{m2}\nonumber\\
&&+2[\frac{4\sqrt{\pi}}{3}\delta_{l 0}\delta_{m0}-\frac{1}{\sqrt{5}}\delta_{l2}\delta_{m0}]\delta(\omega)\}
\end{eqnarray}
\newline\
Thus, the source term (\ref{d7xx}) reads
\begin{eqnarray}\label{qq4}
f_{lm}(r,r_0;\omega;\omega_0)&=&Q_{lm}^{(1)}+Q_{lm}^{(2)}\nonumber\\
&=&-\frac{B_0^2 r_0^6}{r^4}\delta(\omega-2\omega_0)\int e^{2i\phi}Y_{l m}^{*}(\theta,\phi)d\Omega\nonumber\\
&&+\frac{5B_0^2 r_0^6}{r^4}\{\frac{1}{3}\sqrt{\frac{\pi}{5}}\delta(\omega-2\omega_0)\delta_{l 2}\delta_{m2}\nonumber\\
&&+[\frac{4\sqrt{\pi}}{3}\delta_{l 0}\delta_{m0}-\frac{1}{\sqrt{5}}\delta_{l2}\delta_{m0}]\delta(\omega)\}
\end{eqnarray}
Below, we calculate the expression
\begin{eqnarray}\label{ax2}
B_{lm}&\equiv&\int e^{2i\phi} Y_{lm}^{*}(\theta,\phi) d\Omega\nonumber\\
&=&\int_0^{2\pi}e^{-im\phi} e^{2i\phi}d\phi\int_0^{\pi}\sqrt{\frac{(2l+1)}{4\pi}\frac{(l-m)!}{(l+m)!}}P_l^m(\cos{\theta})\sin{\theta}d\theta\nonumber\\
&=&2\pi\delta_{m2}\sqrt{\frac{(2l+1)}{4\pi}\frac{(l-m)!}{(l+m)!}}\int_{-1}^{1} P_l^m(\cos{\theta})d\cos{\theta}
\end{eqnarray}
For simplicity we call
\begin{equation}\label{axf}
M_{lm}\equiv\sqrt{\frac{(2l+1)}{4\pi}\frac{(l-m)!}{(l+m)!}}\int_{-1}^{1} P_l^m(\cos{\theta})d\cos{\theta}
\end{equation}
and thus, eq.~(\ref{ax2}) reads
\begin{equation}\label{bx2}
B_{lm}=2\pi M_{lm}\delta_{m2}
\end{equation}
Eqs.~(\ref{qq4}) and (\ref{bx2}) then  yield
\begin{eqnarray}\label{qq5}
f_{lm}(r,r_0;\omega;\omega_0)&=&-\frac{2\pi B_0^2 r_0^6}{r^4}\delta(\omega-2\omega_0)M_{l m}\delta_{m2}\nonumber\\
&+&\frac{5B_0^2 r_0^6}{r^4}\{\frac{1}{3}\sqrt{\frac{\pi}{5}}\delta(\omega-2\omega_0)\delta_{l 2}\delta_{m2}\nonumber\\
&+&[\frac{4\sqrt{\pi}}{3}\delta_{l 0}\delta_{m0}-\frac{1}{\sqrt{5}}\delta_{l2}\delta_{m0}]\delta(\omega)\}
\end{eqnarray}
%%%%%%%%%%%%%%%%%%%%%%%%%%%%%%%%%%%%%%%%%%%
%%%%%%%%%%%%%%%%%%%%%%%%%%%%%%%%%%%%%%%%%%%%%
In Sec.3, we point out that eq.~(\ref{s21}) is of the form of eq.~(\ref{d5}) and thus, we need to compute the following integrals:
%using the approximate eqs.~(\ref{p13}), namely
First, we compute
\begin{equation}\label{x1}
J_{lm}=\int_{r_0}^{R_{\rm lc}}j_{l}(\omega \rho)f_{lm}(\rho;\omega) d\rho
\end{equation}
%observe that the $f_{lm}$ behaves as $1/r^4$.
Using the approximate eqs.~(\ref{p13}), we find
\begin{eqnarray}\label{x3}
J_{lm}&=&\int_{r_0}^{R_{\rm lc}}j_{l}(\omega r)f_{lm}(r,r_0;\omega,\omega_0) dr\nonumber\\
&=&\int_{r_0}^{R_{\rm lc}}\frac{\omega^l r^l}{(2l+1)!!} f_{lm}(r,r_0;\omega,\omega_0)d r\nonumber\\
&=&-\frac{B_0^2 r_0^6\omega^l}{(2l+1)!!}\frac{R_{\rm lc}^{l-3}-r_0^{l-3}}{l-3}\{2\pi\delta_{m2} M_{lm}\delta(\omega-2\omega_0)\nonumber\\
&&+\frac{\sqrt{5\pi}}{3}\delta(\omega-2\omega_0)\delta_{l 2}\delta_{m2}+\frac{20\sqrt{\pi}}{3}\delta(\omega)\delta_{l0}\delta_{m0}-\sqrt{5}\delta(\omega)\delta_{l2}\delta_{m0}\}\nonumber\\
\end{eqnarray}
Second, using the approximate expressions (\ref{p12pw}), we calculate
\begin{eqnarray}\label{x4}
I_{lm}(r,r_0;\omega,\omega_0)&=&\int_{-\infty}^{\infty}\omega d\omega e^{-i\omega t}H_l^{(1)}(\omega r)J_{lm}(r,r_0;\omega,\omega_0)\nonumber\\
&=&
\frac{iB_0^2 r_0^6}{6\pi r}\frac{2^l\omega_0^l}{(2l+1)!!}\frac{r_{0}^{l-3}-R_{\rm lc}^{l-3}}{l-3}\nonumber\\
&&\cdot\ e^{i[2\omega_0 t-2\omega_0 r+\frac{l\pi}{2}]}[6\pi \delta_{m2} M_{lm}+\sqrt{5\pi}\delta_{l2}\delta_{m2}]\nonumber\\
\label{g4}
\end{eqnarray}
With the aid of eq.~(\ref{d5}) and the eqs.~(\ref{x3}) and (\ref{x4}) we find the solution of the inhomogeneous eq.~(\ref{s21}).


\begin{thebibliography}{}
\bibitem[Arfken et al. 2001]{Arfken2001}  Arfken, G. B., Weber, H. J. 2001, Mathematical Methods for Physicists, Academic Press, San Diego, CA (USA)
\bibitem[Bambi et al. 2022]{Bambi2022} Bambi,C., Katsanevas, S., Kokkotas, K. D. 2022, Handbook of Gravitational Wave Astronomy, Springer
\bibitem[Bogovalov 1999]{Bogovalov1999} Bogovalov, S. 1999, Astron. Astrophys., {\bf 349}, 1017
\bibitem[Contopoulos et al. 2022]{Contopoulos2022} Contopoulos, I., Strantzalis, A., Papadopoulos, D. B., Kazanas, D. 2022, Mon. Not. R. Astr. Soc., {\bf 509}, 174
\bibitem[Ivanov 2008]{Ivanov2008} Ivanov, B. V. 2008, Phys.Rev.D., {\bf 77}, 044007
\bibitem[Jackson 1999]{Jackson1999} Jackson, J. D. 1999, Classical Electrodynamics, John Wiley and Sons, Inc., third Ed.
\bibitem{r5} Korn, G. A., Korn, T. M. 1968, Mathematical Handbook for Scientists and Engineers: Definitions, Theorems and Formulas for Reference and Review (MacGraw-Hill Book Co, (NY)
\bibitem[Maggiore 2008]{Maggiore2008} Maggiore, M. 2008, Gravitational Waves, Oxford University Press
\bibitem[Mastrano et al. 2015]{Mastranoetal2015} Mastrano, A., Suvorov, A. G., Melatos, A. 2015, \mnras, {\bf 447}, 3475
\bibitem[Misner et al. 1973]{Misner1973} Misner, C. W., Thorne, K., Wheller, J. 1973, Gravitation, W.H.Freeman, San Francisco
\bibitem[Mpore et al. 2014]{Mooreetal2014}Moore C. J., Cole R. H., Berry C. P. 2014, CQG, {\bf 32}, 015014
\bibitem[Muslimov \& Harding 2009]{Muslimov2009} Muslimov, A. G., Harding, A. K. 2009, Astrophys. J., {\bf 692}, 140
\bibitem[Nazari \& ROshan 2020]{Nazarietal2020} Nazari, E., Roshan, M. 2020, \mnras, {\bf 498}, 110
\bibitem[Persides 1974]{Persides1974} Persides, S. 1974, J. Math. Phys., {\bf 15}, 885
\bibitem[Rainich 1925]{Rainich1925} Rainich, G. Y. 1925, Trans. Amer. Math. Soc., {\bf 27}, 106
\bibitem[Thorne et al. 1982]{Thorne1982}  Thorne, K., Macdonald, D. 1982, \mnras, {\bf 198}, 339
\bibitem[Turyshev et al. 2018]{Turyshev2018} Turyshev, S. G., Toth, V. T. 2018, arXiv:1801.06253 [physics.optics]
\bibitem[Wald 1984]{Wald1984} Wald, R. M. 1984, General relativity, The University of Chicago, Ltd., London
%\bibitem{Thorne1982} C.W. Misner and Wheller, Ann. Phys., {\bf 2}, 525 (1957)
%%%%%%%%%%%%%%%%%%%%%%%%%%%%%%%%%%%%%%%%%%%%%%%%%%%%%%%%%%%%%%%%%%%%%%%%%%%%%%%%%
%\bibitem{r3} Masashi Kimura et al, arXiv:2105.05581 [gr-qc] (2021)
%\bibitem{r4} Akash Patel et al, arXiv: 2108.01788  [gr-qc] (2021)
%%%%%%%%%%%%%%%%%%%%%%%%%%%%%%%%%%%%%%%%%%%%%%%%%%%%%%%%%%%%%%%%%%%%%%%%%%%%%%%%%%%%%
%\bibitem{a6p}R.Adler,M.Bazin,M. Schiffer, Introduction to General Relativity,McGraw-Hill,Ny, pag. 343 (1975)
%%%%%%%%%%%%%%%%%%%%%%%%%%%%%%%%%%%%%%%%%%%%%%%%%%%%%%%%%%%%%%%%%%%%%%%%%%%%%%%
%\bibitem{r5} G.A.Korn and T.M.Korn, Mathematical Handbook for Scientists and Engineers:Definitions,Theorems, and Formulas for Reference and Review (MacGraw-Hill Book     Co,NY), (1968)
%\bibitem{Turyshev2018} Slava G. Turyshev, Victor T. Toth, arXiv:1801.06253 [physics.optics] (2018)
\end{thebibliography}
\end{document}